\newcommand{\FtD}{\tilde{F}_2^D(\beta,Q^2)}

\newcommand{\flp}[1]{$ {F_2^{LP\,( #1)}}$}

\newcommand{\do}{D^0}

\newcommand{\pom}{I\!\!P}

\newcommand{\xpom}{x_{\pom}}

\newcommand{\x}{\mbox{$x$}}

\newcommand{\MeV}{\mathrm{MeV}}

\newcommand{\GeV}{\mbox{\rm GeV}}

\newcommand{\gevsq}{\mbox{${\rm GeV}^2$}}

\newcommand{\pbinv}{\mbox{${\rm pb^{-1}}$}}

\newcommand{\md}{\mbox{d}}

\newcommand{\hdick}{\noalign{\hrule height1.4pt}}

\documentstyle[epsfig]{aipproc}

\begin{document}
\title{H1 Results in Deep Inelastic Scattering}

\author{
Gregorio Bernardi} 
\address{ {\rm H1 Collaboration } \\ 
LPNHE -- Universit\'e de Paris 6-7,   CNRS-IN2P3 \\
{\it e-mail: gregorio@mail.desy.de} }    
\maketitle

\vspace*{-0.5cm}
\begin{abstract}
Recent deep inelastic results from the H1 collaboration are presented.
The topics covered include inclusive  diffractive physics,
the determination of $\alpha_S$ using the event shapes,
the study  of the hadronic final state at low $x$  with
single particles and with jets, the 
structure function measurements in particular $F_2^p$ at low $Q^2$ and
the high $Q^2$ physics both in neutral and charged current.
\end{abstract}

\vspace*{-0.5cm}
\section{ { {Introduction}}}
The first two years after the commissionning of the first $ep$ collider
HERA in May 1992 were devoted in deep inelastic scattering (DIS) to
the observation of the new processes which would become the main subject
of study nowadays in DESY. The observation of the rise of the $F_2^p$
proton structure function  at low $x$~\cite{bb.h1f292,bb.zef292} was the first
of this type of 
measurement which is still progressing in terms of precision and 
possibilities of QCD test, $\alpha_S$ and gluon density determination.
The rich phenomenology at low $x$ could also be studied 
in the measurement of hadronic final state properties.
The clear observation of diffractive  events~\cite{bb.zediff92,bb.h1diff92} 
paved the way to a detailed and not yet finished 
study of one of the most obscure problem
of the strong interaction. The observation of high $Q^2$ charged and neutral
current~\cite{bb.h1cc92,bb.zecc92}
 with very low statistics at that time, but in a perfect experimental
environment allowed a glimpse to what will be the future of the HERA collider.
The subsequent years (1994,1995,1996) are  covered in this report.
They allowed the first precise measurements of the phenomena mentioned
above, besides many others. In 1995, the H1 collaboration upgraded
the backward detectors of the experiment, introducing in particular
a more precise drift chamber ({\small BDC}) to measure the polar angle of the scattered
electron, and a new {\small SPACAL}
calorimeter with better hadronic containment,
better granularity and better angular acceptance. The results obtained
have fullfilled the expectations and a new step in precision could thus be
reached for the low $Q^2$ physics.

\begin{figure}  
  \begin{center}
   \epsfig{file=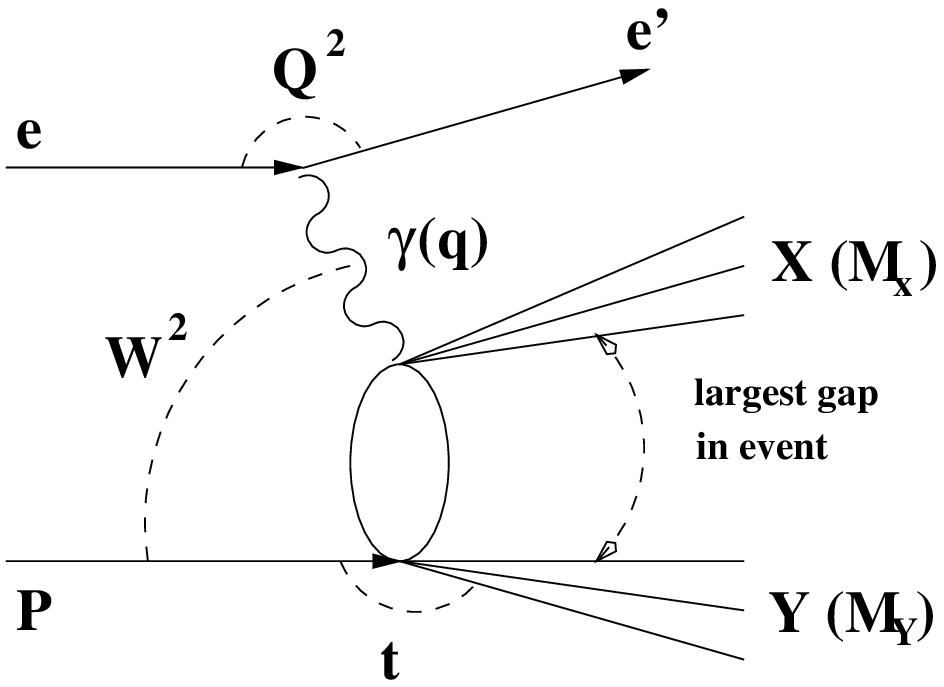,width=7.cm}
\label{myfig.1}
  \end{center}
\end{figure}
\vspace*{-2.cm}
{\small {\bf FIGURE 1:}}
\vspace*{0.2cm}
\begin{center}
\begin{tabular}{ll}
{\bf Inclusive Variables}\hspace*{2cm} & {\bf Diffractive Variables}\\
$Q^2 = -q^2$ & $ \x =\beta \cdot \xpom$\\
$ x = Q^2 / (2 \ q \cdot P)$ &  $\beta~=~{Q^2}/({2 \ q\cdot(P-Y)})
            \hspace*{0.4cm}        \simeq~\frac{Q^2}{Q^2+M_X^2}$ \\
$ y = (q \cdot P) / (e \cdot P)$  & \\
$ W = Q^2\ (1-x) / x \simeq y s $  &
$\xpom~=~{q\cdot (P-Y)}/({q\cdot P})~
              \simeq~\frac{Q^2+M_X^2}{Q^2+W^2}$  \\
$ M = \sqrt{sx} = \sqrt{Q^2 / y} $ & $ t=(P-Y)^2$ \\
 & \\
\end{tabular}
\end{center}
In the following we will often refer to the inclusive and
diffractive kinematic variables, and to their method of reconstruction so
let us introduce them here, using fig.~1
as an illustration of the basic
DIS process. The exchanged boson is a photon in most of the cases, but
the exchange of $Z^{\circ}$ and $W^{\pm}$  becomes sizeable at
high $Q^2$. 
The basic DIS selection requires the detection
of an electromagnetic cluster representing the scattered electron\footnote
{HERA can run both with $e^+$ and $e^-$, but here ``electron''
is used both for $e^+$ and $e^-$.} above a certain energy threshold which is
typically around 10 GeV, and the presence of a reconstructed interaction vertex
in order to reject the beam associated background and to improve the 
reconstruction quality. Further requirements are applied 
depending on the analyses as we will see later. The hadronic final state is
resolved in two systems X and Y in the study of diffraction using
a specific selection.

The kinematics can be reconstructed at HERA with
different  methods for the  neutral current (NC) events (electron only
($e$), $\Sigma$, Double Angle (DA)), but only one (hadrons only, $h$)
for the
charged current (CC) 
since then only the hadronic final state is measured.
In general H1 uses for the NC
the $e$ and the $\Sigma$ method, since they complete
each other. The $h$ method is used for the CC analysis,
while the DA has been used as a cross-check in the high $Q^2$ NC
analysis.

Section II discusses the  diffractive physics results, section III 
the determination of $\alpha_S$ using the event shapes,
section IV
the study  of the hadronic final state at low $x$  with
single particles and with jets, section V the 
structure function measurements and section VI
the high $Q^2$ NC and CC physics. 

\section{ Inclusive Diffraction}
The DIS diffractive events are characterized by a  hadronic final state having
a large gap in
rapidity (see fig.~1)
between  the system X 
 which is associated
to the interacting parton and the system Y which is related to the
fragments of the proton. The interaction can be described by
the exchange of a colourless object (a ``pomeron'', $\pom$)
which creates the gap observed between
the two systems. Experimentally these DIS events are selected by the
absence of hadrons in the pseudorapidity interval $3.4 < \eta < 7.5$.
The system X (Y) is composed of all particles produced backward\footnote{The 
positive $z$ axis is defined at HERA as the incident proton beam direction.}
(forward) of this gap. The limit $\eta=7.5$ for the system Y restricts
its mass to M$_Y < 1.6$ GeV and its squared momentum to $|t| < 1.0$ GeV$^2$.
Hard diffraction can be quantified by the measurement of the
diffractive structure function
 $F_2^{D(3)}$ which  is related to the triple differential cross-section
of diffractive events via:
\begin{displaymath}
\frac{\md^3 \sigma^D_{ep \rightarrow e^{\prime} XY}}{\md\beta \md Q^2 \md
\xpom} =
\frac{4\pi\alpha^2}{\beta Q^4}
(1-y+\frac{y^2}{2}) \cdot F_2^{D(3)}(Q^2,\beta,\xpom)
\end{displaymath}
The kinematic range of the  1994 measurement is 
 $2.5<    Q^2    <65\,\GeV^2$,   
 $0.01<    \beta  <0.9  $  and $0.0001<  \xpom  <0.05 $.
This analysis  was already reported by H1~\cite{bb.andy_eilat}, but
the interpretation of the measurement has progressed since that 
first observation
of the ``factorization breaking''.
This ``breaking'' means that in $F_2^{D(3)}$ a
pomeron flux of the type $\xpom^{-n}$ cannot be factorized out,
unless $n$ is taken as a function of $Q^2$ and $\beta$.  H1 has
already shown with the 1994 data that  $n$ depends on $\beta$
but not on $Q^2$~\cite{bb.julian_warsaw}. In fig.~2
 is shown the quantity
$\xpom \cdot F_2^{D(3)}(Q^2,\beta,\xpom)$ as a function of $\xpom$
for different values of $\beta$ and $Q^2$. The deviations from
factorization are clearly visible and a possible explanation 
is the  contribution from so-called sub-leading trajectories
to the measured cross-sections which may be identified with the
exchange of particles carrying the quantum numbers of the physical meson
states. After the pomeron, the next-leading exchanges are the
trajectories of the $f_2,$ $\rho$, $\omega$ and $a_2$ mesons, which are
expected to give a contribution to $\xpom \cdot F_2^{D(3)}(Q^2,\beta,\xpom)$ 
which rises with $x$.

In fig.~2 is also shown the result of a fit to the data in which 
both a pomeron component of the type A$(\beta,Q^2) \cdot \xpom^{-n_{\pom}}$
and a meson component $C_M \cdot F_2^M(\beta,Q^2) \cdot \xpom^{-n_M}$
contribute to $F_2^{D(3)}$ (see \cite{bb.julian_warsaw} for details)
The parameters $C_M$, $n_{\pom}$ and $n_M$ are free parameters
in the fit together with  $A(\beta,Q^2)$ at each value of $\beta$ and $Q^2$.
The function $F_2^M(\beta,Q^2)$ is taken from the GRV parametrization 
of the pion structure function~\cite{bb.grv-pion}.
The result of the
fit assuming a maximal interference between the pomeron and meson amplitudes
is shown in the upper curve of each bin of fig.~2,
 and has a $\chi^2/ndf=
165/156$. The lower line represents the pomeron contribution alone, while
the middle one represents the sum of the two contribution without taking into
account the interference. The influence of the meson exchange is limited 
at relatively high $\xpom$ ($>0.05$) and is stronger at low $\beta$.
After correcting for the integration over $t$, the resulting pomeron
and meson intercepts 
are 
\begin{center}
 $
\alpha_{\pom}(0)=1.18\pm 0.02\mbox{(stat)}\pm 0.04\mbox{(syst)} 
\hspace*{0.5cm} ;  \hspace*{0.5cm} 
 \alpha_{M}(0)=0.6\pm 0.1\mbox{(stat)}\pm 0.3\mbox{(syst)} $ 
\end{center}
While the value  of $\alpha_M$ is consistent with that expected for the
exchange of the mesons previously cited, the value of  $\alpha_{\pom}(0)$
is somewhat higher than the one obtained when parametrizing the total
cross-section of soft hadronic interaction~\cite{bb.dola}.  

A QCD analysis~\cite{bb.dirkmann} of $\FtD$, which is obtained by integrating  
$F_2^{D(3)}(Q^2,\beta,\xpom)$ between $0.0003 < \xpom < 0.05$,
reveals the partonic structure of the pomeron (mesonic effects
are here neglected, but have been checked not to change the interpretation).
It is found to be dominated by a hard gluon density peaking at $x_{g/{\pom}}
\simeq 1$, and about 80\% of its momentum is carried by gluons at $Q^2$ below
65 GeV$^2$~. This result is confirmed by the analysis of the hadronic final 
state of DIS diffractive events~\cite{bb.cormack}.
\vspace*{-1cm}
\begin{figure}[h]
\epsfig{file=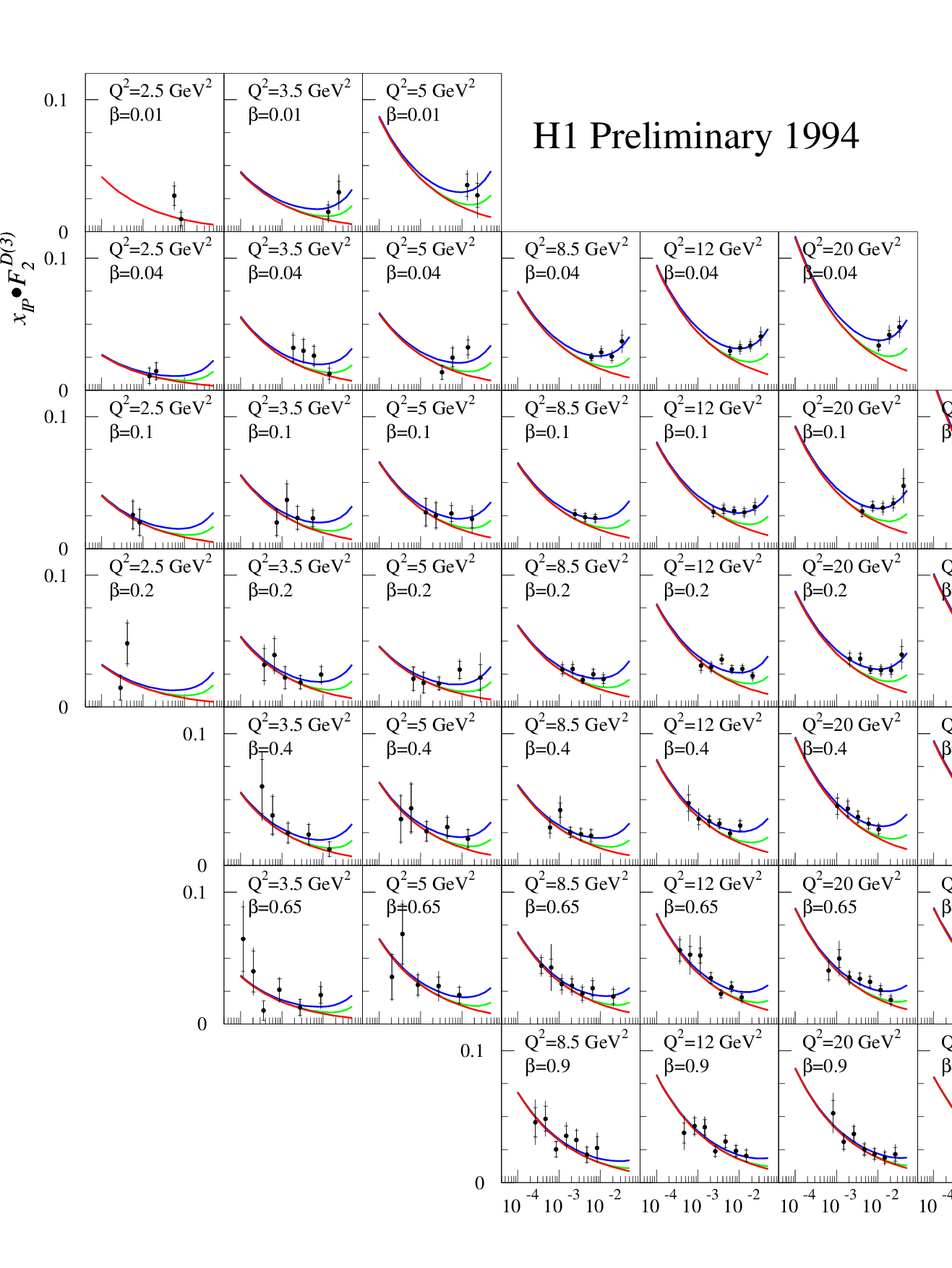,bbllx=0,bblly=20,bburx=650,bbury=900,
width=11cm,,height=13.cm,clip=}
\label{myf2d3}
\end{figure}

\noindent
{\small{\bf FIGURE 2.}}
{\small {\sl
$\xpom \cdot F_2^{D(3)}(Q^2,\beta,\xpom)$ as a function
of $\xpom$ for different values of $\beta$ and $Q^2$. 
Also
shown is the phenomenological fit described in the text.}}
\begin{center}
{\large {\bf{Diffractive DIS with a Leading Proton in the \\
  Forward Proton Spectrometer (FPS)}}}
\end{center}

In 1995 the FPS has been installed in H1 allowing  leading protons
with energies between 500 and 760 GeV and scattering angles below 1 mrad
to be detected. Its main use is, after  detection of a  leading proton, 
to measure the variable
 $\xi=1-E'_p/E_p$, where  $E_p$ and $E'_p$ are the
proton energy before and after the interaction. $\xi$
is defined as $\xpom$ in terms of 4-vectors and may be
interpreted as the fraction $x_{\pi/p}$ of the proton's 4-momentum carried
by the exchanged object. It is thus possible to
 define a Leading Proton structure function \flp{3} in the same way as
$F_2^{D(3)}$, but in which 
$\beta$ is interpreted  as the fraction $x_{q/\pi}$ of the
momentum of the particle $\pi$ carried by the struck quark. 
 The structure function 
\flp{3} shown in fig.~3 has been measured for events with
  a proton of  $p_\perp < 200\,\MeV$ and
 $E_p' = 580 - 740\,\GeV$ using
the data collected in 1995 which represent an integrated luminosity of 
1.44 pb$^{-1}$~\cite{bb.blist}.

 The measurement  shows a weak $\xi$ dependence as expected for
pion exchange and a logarithmic rise with $Q^2$. Although
the behaviour is reproduced in shape by a pion-exchange model introduced
in RAPGAP~\cite{bb.rapgap} (GRV-LO pion parton densities are used), the overall
normalization of the simulation is too low by about a factor of two.
The factorization of \flp{3} as $f_{\pi/p}(\xi) \cdot F_2^{\pi}(\beta,Q^2)$
could not be established yet, although the data are not incompatible with this
hypothesis. The  upgraded FPS   and the additional data taken in 1996 
should allow to answer these questions in the near future.
\vspace*{-1.2cm}
\begin{figure}[h]
\begin{center}
{\epsfig{file=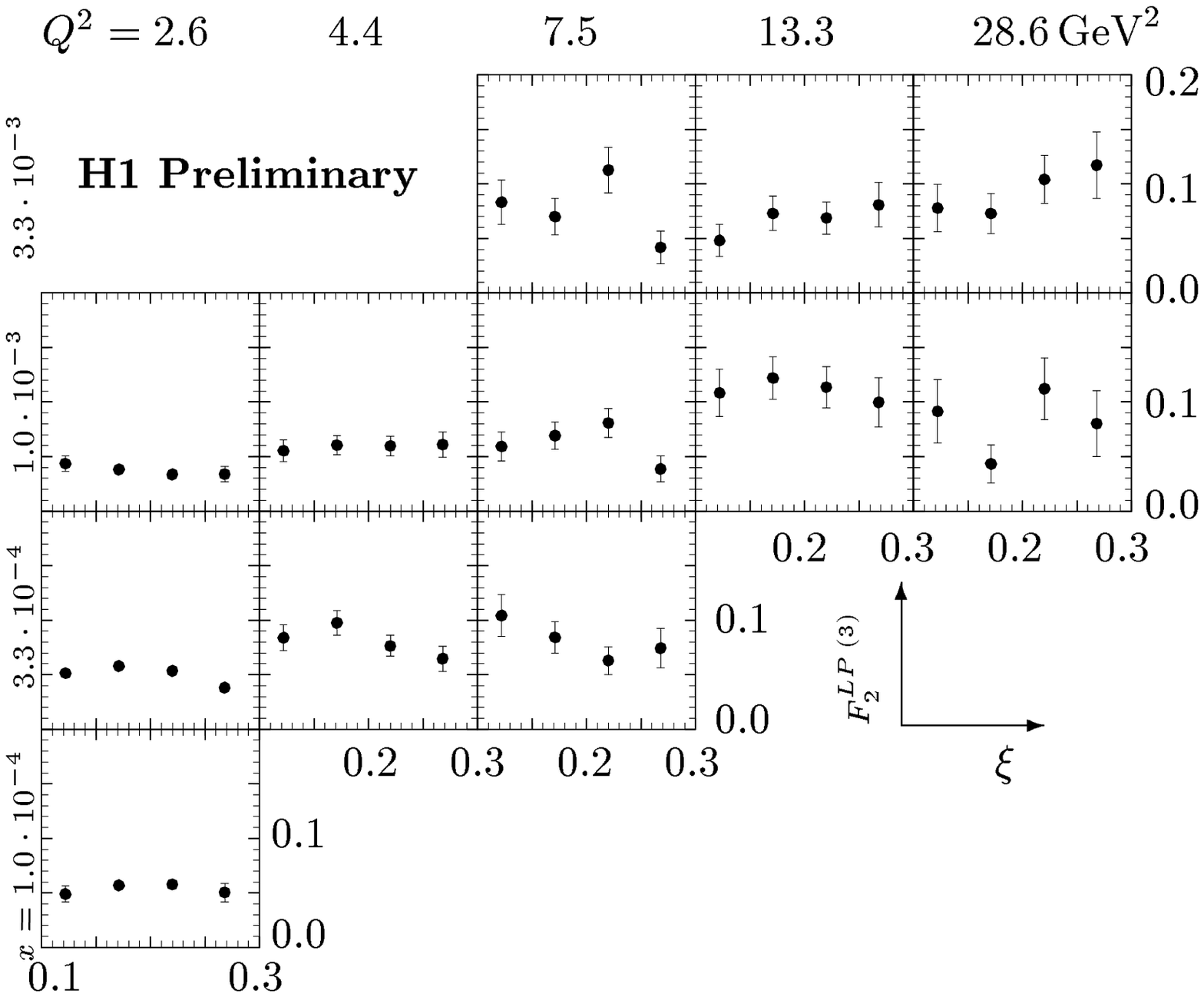,width=12cm,height=9.cm,
bbllx=25,bblly=260,bburx=510,bbury=640}}
\label{myflp3} 
\end{center}
\end{figure}

\vspace*{-0.3cm}
\noindent
{\small {\bf FIGURE 3.}}
{\small \sl {Measurements of \flp{3} as a function of $\xi$ in bins of $x$ and
$Q^2$.
}}

\setcounter{figure}{3}

\section{ $\alpha_S$ from the DIS Event Shapes}
At HERA,
 the DIS event shapes
can be studied in the ``current'' hemisphere of 
 the Breit frame\footnote{
Defined such that the exchanged boson is purely space-like with 4-momentum
(0,0,0,-Q).}
which is analogous to one of the two hemispheres measured at 
 LEP \cite{bb.h1es}.
The scale correponding to  $\sqrt{s}/2$ in $e^+
e^-$ collisions is $Q/2$, i.e. at HERA it is possible to vary continuously
this scale and study the evolution of the observables as a function of $Q$.
The event shapes variables which have been studied are 
the jet mass $\rho_c$, the jet broadening ($B_c$),
the thrust $T_c$ and
$T_z$ i.e.
respectively  defined 
along the thrust axis (${\bf n_T}$) or along the hemisphere axis (${\bf n}$),
 We refer 
to~\cite{bb.rabbertz_chigago} 
for a discussion of $B_c$ and concentrate
here on the other 3 variables which are defined as
$$
       T_c  =  \max \, \frac{\sum_h |\, {\bf p}_h\cdot {\bf n}_T \, |} 
                 {\sum_h |\, {\bf p}_h \, |} 
\quad
      T_z  =  \frac{\sum_h |\, {\bf p}_h\cdot {\bf n} \, |} 
                 {\sum_h |\, {\bf p}_h \, |} 
\quad
      \rho_c  =  \frac{M^2}{Q^2} 
      \ = \ \frac{(\, \sum_h \, p_h \, )^2}{Q^2} 
$$
The sums extend over all hadrons $h$ with 4-momentum $p_h=[E_h,{\bf p}_h]$
fullfilling $\cos{({\bf p}_h \cdot {\bf n})}>0$, where the current hemisphere
axis coincides with the exchanged boson direction.
The study has been made  for  8 GeV $< Q <$ 68 GeV using 2.9
pb$^{-1}$ of 1994 data at low
$Q^2$ and 10.9 pb$^{-1}$ in the 1994-1996 data at high $Q^2$.

\begin{figure} \centering 
\epsfig{file=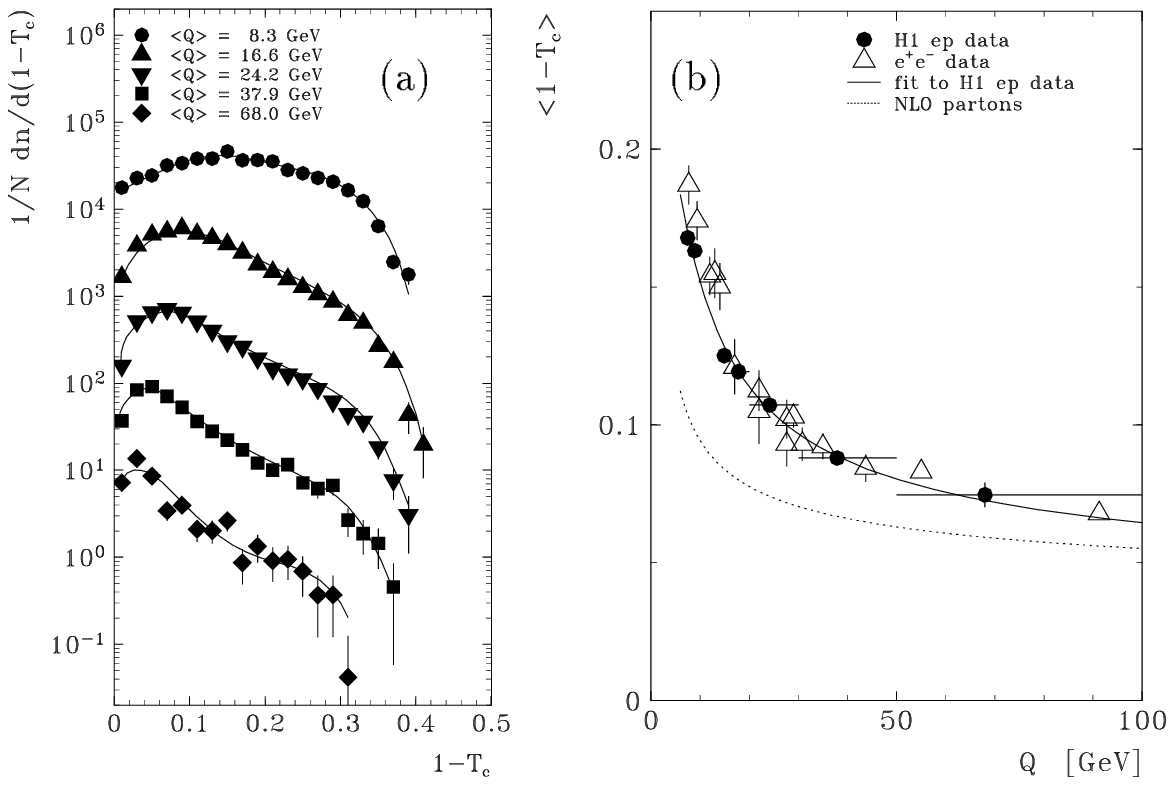,width=12cm,
bbllx=82.,bblly=342.,bburx=425.,bbury=565.}
\caption[]{\label {fig.meanTc}     
{\small \sl 
a) Distributions of 1-$T_c$ in bins of $Q^2$.
b) Comparison of the evolution of the mean of 1$-T_c$ as a function of $Q$
in DIS and $e^+e^-$.}}
\end{figure}

The normalized differential spectrum of the $T_c$ variable and its mean
value as a function of $Q$ are  shown as an example 
in fig~\ref{fig.meanTc}. The mean value exhibit a strong $Q$ dependence,
decreasing with rising $Q^2$, i.e. the energy flow in the current
hemisphere becomes more collimated.
The data are 
well described by the LEPTO Monte Carlo model~\cite{bb.lepto}
 for all $Q$. This distribution
is in gross agreement with the $e^+e^-$ data. The small differences can
be understood  as due to the different analysis methods used at HERA and LEP,
and to the different physics effects related to the nature of the interaction 
involved. 

The QCD analysis of these distributions is
built upon the fact that the mean value of any ``infrared safe''
event shape variable $F$ such as $1-T_c, (1-T_z)/2$ or $\rho_c$
can be decomposed   in DIS  and in 
    $e^+e^-$ annihilation as~\cite{bb.doksh}
    \begin{eqnarray*}
      \langle F \rangle  = 
      \langle F \rangle^{{\rm pert}} + 
      \langle F \rangle^{{\rm pow}}  
    \end{eqnarray*}
with the perturbative       $ \langle F \rangle^{{\rm pert}} $
and the power correction part $      \langle F \rangle^{{\rm pow}}  $
given by 
    \begin{eqnarray*}
      \langle F \rangle^{{\rm pert}} 
        & = & c_1\,\alpha_s(Q) +  c_2\, \alpha^2_s(\mu_R) \\
      \langle F \rangle^{{\rm pow}} & = & 
      a_F\,\frac{16}{3\,\pi}\,\frac{\mu_I}{Q}\, 
      \ln^p \frac{Q}{\mu_I}\,
      \left [ \,
      \bar{\alpha}_0(\mu_I)
      -\, \alpha_s(Q) 
      - \frac{\beta_0}{2\,\pi}\, 
      \left ( \ln\frac{Q}{\mu_I} + \frac{K}{\beta_0} + 1 \right ) 
      \alpha^2_s(Q) \, \right ] 
    \end{eqnarray*}
The  coefficients $c_1,\ c_2$ are obtained from  
${\cal O}(\alpha_s^2)$ DISENT calculations~\cite{bb.disent}. The power
(or hadronization) corrections are believed to stem from a universal soft gluon
phenomenon associated with the behaviour of the 
running coupling at small momentum scales, i.e. 
the  usual   $1/Q$ corrections  are not necessarily 

\vspace*{-0.5cm}
\noindent
related to hadronisation.
The universality means that the phenomenon could be described by  
calculable        coefficients {$a_F$} and {$p$} which are  process dependent
together with
a few  non-perturbative parameters like  $\bar{\alpha}_0(\mu_I)$
which have       to be evaluated at some `infrared matching' scale
 $\Lambda \ll \mu_I \ll Q$.
The values of the means at the different $Q^2$ values can thus be fitted
to give a simultaneous determination of  $\bar{\alpha}_0(\mu_I)$ and
of the strong coupling constant $\alpha_S(M_Z)$ independently of any
fragmentation model. The results are given below for the H1 $ep$ data, followed
by the parameters obtained in  $e^+e^-$ data for comparison.
\begin{center}
    \begin{tabular}{l c c c}
      Observable     &
        $\bar{\alpha}_0(\mu_I = 2~\GeV)$ & $\alpha_s(M_Z)$ &
        \ $\chi^2/\mbox{ndf}$ \
      \\[1ex] \hdick \\[-1.5ex]
      $\langle 1 - T_c \rangle$ &
        $ \ 0.497 \pm 0.005 \ ^{+0.070}_{-0.036} \ $ &
        $ \ 0.123 \pm 0.002 \ ^{+0.007}_{-0.005} \ $ &
        $ \ 5.0/5 \ $ 
      \\[.4ex]  
      $\langle 1 - T_z \rangle \, / \, 2$ &
        $ \ 0.507 \pm 0.008 \ ^{+0.109}_{-0.051} \ $ &
        $ \ 0.115 \pm 0.002 \ ^{+0.007}_{-0.005} \ $ &
        $ \ 8.5/5 \ $ 
      \\[.4ex]  
      $\langle \rho_c \rangle$ &
        $ \ 0.519 \pm 0.009 \ ^{+0.025}_{-0.020} \ $ &
        $ \ 0.130 \pm 0.003 \ ^{+0.007}_{-0.005} \ $ &
        $ \ 3.1/5 \ $ 
      \\[.4ex]
       common fit & 
        $ \ 0.491 \pm 0.003 \ ^{+0.079}_{-0.042} \ $  &
        $ \ 0.118 \pm 0.001 \ ^{+0.007}_{-0.006} \ $  &
        $ \ 39/19 \ $ \
      \\[1ex] \hdick \\[-1.5ex]
        $\langle 1 - T_{ee} \rangle$ &
        $ \ 0.519 \pm 0.009 \ ^{+0.093}_{-0.039} \ $ &
        $ \ 0.123 \pm 0.001 \ ^{+0.007}_{-0.004} \ $ &
        $ \ 10.9/14 \ $ 
      \\[.4ex]  
      $\langle M^2_H/s \rangle$ &
        $ \ 0.580 \pm 0.015 \ ^{+0.130}_{-0.053} \ $ &
        $ \ 0.119 \pm 0.001 \ ^{+0.004}_{-0.003} \ $ &
        $ \ 10.9/14 \ $ 
      \\[1ex] \hdick \\[-1.5ex]
  \end{tabular}
\end{center}
The event shapes are compatible with a universal power correction 
parameter {$\bar{\alpha_0} \approx 0.5$} within {$\pm 20\,\%$} which is
also valid at LEP. The $\alpha_s(M_Z)$ determination has the same
precision than the one obtained from a large set of event shape
analyses in $e^+e^-$ at the $Z$ resonance~\cite{bb.13-delphi}, and the
error is here  dominated by theoretical uncertainties due to 
as yet unknown higher order QCD corrections.

\section{BFKL effects in the Hadronic Final State}
\begin{figure}[htbp]
  \begin{center}
\vspace*{-0.5cm}
    \epsfig{file=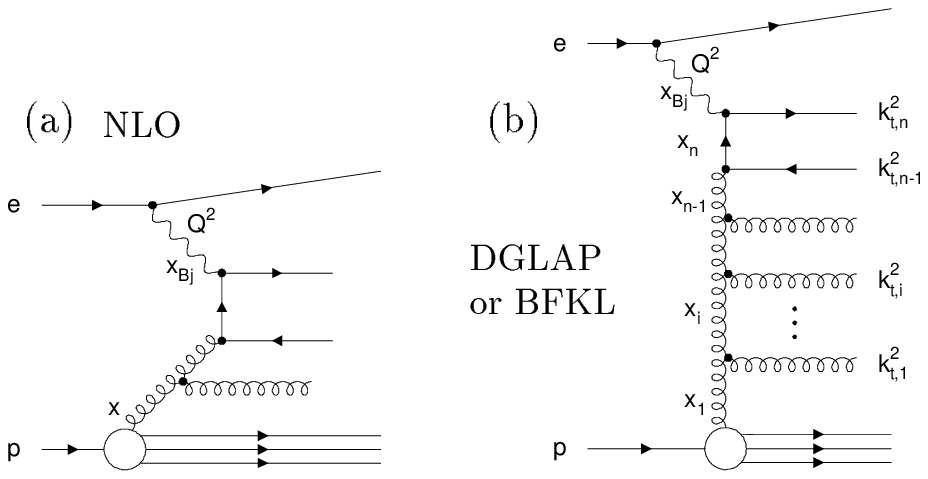,width=10cm,
bbllx=95.,bblly=610.,bburx=370.,bbury=750.}
\caption[]{\label{mydglap_nlo}      
{\small \sl 
a) Schematic representation of DIS in NLO and \\ b) in a parton cascade model
of the DGLAP or BFKL type.}}
  \end{center}
\end{figure}

At HERA the hadronic final state is measured in a precise way,
and renders  possible a test of the QCD dynamics  at high $p_T$.
The analyses described below are performed at $Q^2$ between 5 and 100 GeV$^2$
and between $10^{-4}$ and  $10^{-2}$ in $x$. The measured distributions
are corrected for detector effects to the hadron level,
and compared to   ${\cal O}(\alpha_s^2)$  next to leading
order (NLO) calculations or to 
models based on  QCD phenomenology, see fig.~\ref{mydglap_nlo}. 
 For the dijet 
production rate these NLO calculations
are available and implemented in the DISENT~\cite{bb.disent} and 
MEPJET~\cite{bb.mepjet} Monte Carlo programs. 

Another approach is
to perform a resummation of the leading logarithms of the DGLAP~\cite{bb.dglap}
type ($(\alpha_s \ln \frac{Q^2}{Q^2_0})^n$ terms) or of the BFKL~\cite{bb.bfkl}
type $(\alpha_s \ln \frac{1}{x})^n$ terms) the latter being expected
to dominate at low $x$. The DGLAP approach is used in the LEPTO~\cite{bb.lepto}
and HERWIG~\cite{bb.herwig} event generators in which leading log partons
showers are added to leading order matrix elements (MEPS) in combination
with a string or cluster hadron fragmentation model. 
The BFKL resummation  is not yet available in event generators,
but since there is no $k_t$ ordering in this approach, contrarily to the DGLAP
one, and since this characteristic is also present in the colour dipole
model (CDM) as implemented in the ARIADNE~\cite{bb.ariadne} event generator,
it is possible to have first clues on the
underlying dynamics by confronting the data to ARIADNE. 

\begin{figure}[htbp]
  \begin{center}
\epsfig{file=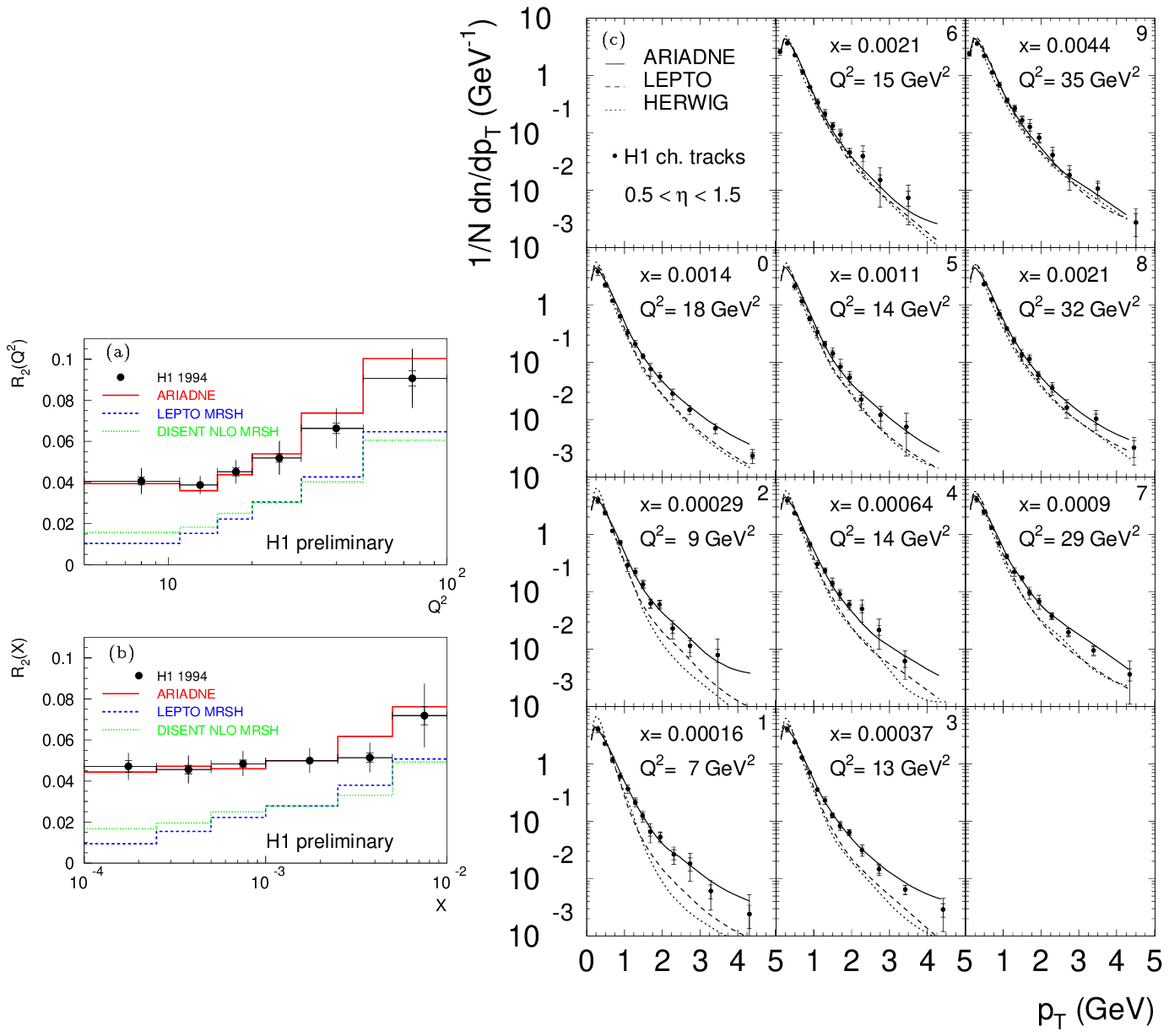,height=10.cm,
bbllx=82.,bblly=422.,bburx=480.,bbury=775.,clip=}
\caption[]{\label{myspectra}      
{\small \sl 
Dijet rate as a function of $Q^2$ (a) and $x$ (b).
c) Distribution of the transverse momentum of central charged
tracks in bins of $Q^2$ and $x$.}}
\end{center}
\end{figure}

To measure the dijet rate,
events with two jets having a cone radius of one and a transverse
energy $E_T>5$ GeV are selected if the rapidity difference between the jets in 
the hadronic center
of mass (h.c.m.) is: $\Delta
\eta^*_{max} < 2$.  The dijet rate, given by the ratio of two-jets
events to the total number of DIS events in the same kinematical region,
is presented in fig.~\ref{myspectra}a,b as a function of $Q^2$ and $x$
and compared to the QCD models and calculations predictions. 
The description of the data by
ARIADNE is excellent but this cannot be interpreted  yet as a proof
for a BFKL effect.
Surprisingly,
LEPTO fails to describe the data even at high $Q^2$ and  moderate $x$,
i.e. in a range where the DGLAP models should be applicable.
This failure cannot be ascribed to the hadronization correction
which are expected to be  relatively small ($<25\%$).
Indeed, this  can be verified  in fig~.~\ref{myspectra}a,b by the good
agreement between  DISENT and LEPTO. 

Recently a sensitive test of BFKL evolution based on the
transverse momentum spectra  of single particles has been 
proposed~\cite{bb.kuhlen}. The hard tail of the $p_T$ spectrum 
was shown to originate from parton radiation while hadronization
effects were suppressed.  Since parton emissions in the central region
of the h.c.m. are less restricted for a $k_T$ unordered scenario,
a harder $p_T$ spectrum than those expected in DGLAP 
models  could be a BFKL signature. 

The measured charged particle spectra~\cite{bb.h1spectra} 
(for 0.5$< \eta^* < 1.5$) are shown in fig.~\ref{myspectra}c.
While at large values of $x$ and $Q^2$ all models describe the data well,
the DGLAP based models fail to describe the high $p_T$ tail at small $x$.
ARIADNE gives a good description of the data over the full kinematic range,
hinting that the first BFKL effects could be in sight.

\clearpage
\newpage


\section{Structure Functions}

\begin{figure}[htb]
\begin{center}
\epsfig{file=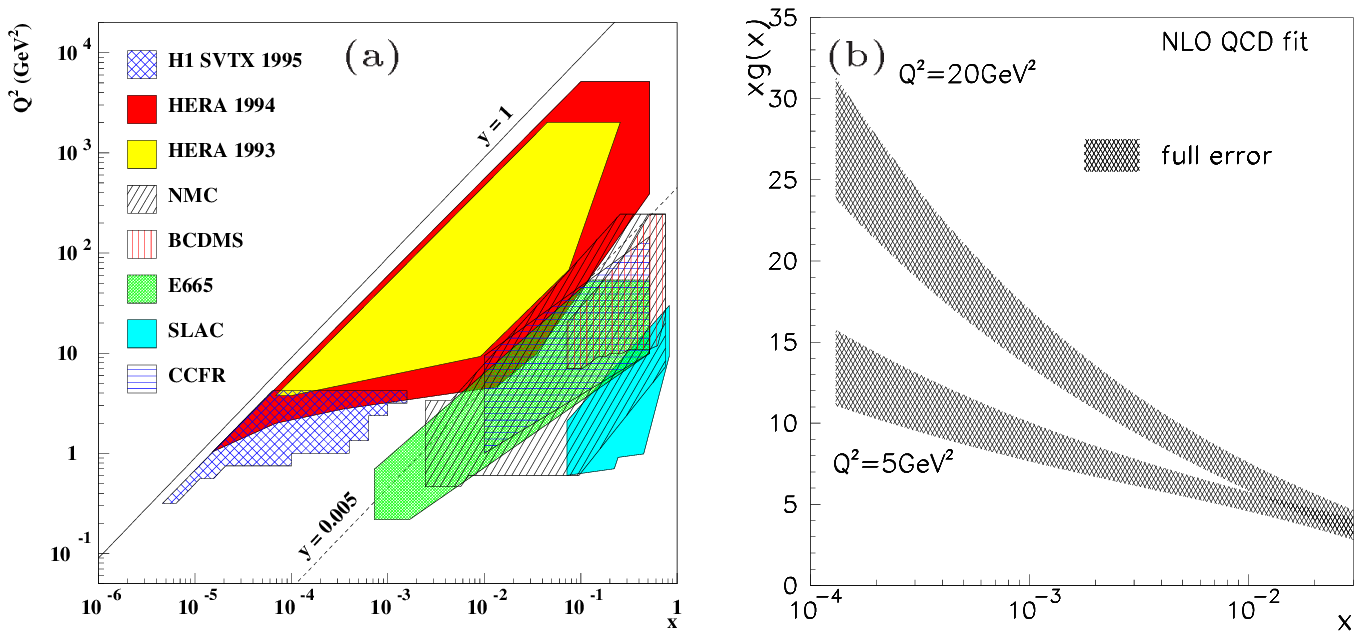,bbllx=115,bblly=595,bburx=495,bbury=775,
width=12.5cm}
\caption[]{\label{fig.kin_h1}      
{\small \sl 
a) The kinematic region of various structure function measurements.
b) The gluon density at $Q^2 = 5$ and $20$ GeV$^2$ extracted by
 H1 NLO-DGLAP fit.
}}
\end{center}
\end{figure}

One of the major physics issues at HERA is 
the measurement of the proton structure functions $F_2^p(x,Q^2)$,
$F_L^p(x,Q^2)$ and $xF_3^p(x,Q^2)$, this last one 
being not  measured yet  at HERA due to lack of statistics.
The rise of $F_2$ with decreasing $x$ has already been established
from the first measurements with the data taken in 1992 and 1993 
\cite{bb.h1f292,bb.zef292,bb.h1f293,bb.zef293}.
Since then, impressive progress has been made, both in the extension
of the kinematic range of the $F_2$ measurement as illustrated in 
fig.~\ref{fig.kin_h1}a, and in the precision achieved. 
\vspace*{-0.8cm}

\vspace*{-0.8cm}
With a luminosity of about 2 $\pbinv$ accumulated in the 94 running
period $F_2^p(x,Q^2)$ has been measured by 
H1~\cite{bb.h1f294,bb.zef294} 
between $1.5-5000~\gevsq$ in $Q^2$ and $0.78-0.01$ in $y$, 
reaching at low $y$ the kinematic region of the fixed target experiments.
In the high statistic region an error of about $5\%$ was achieved, 
allowing precise QCD tests. A NLO-DGLAP fit was performed 
on this data (restricted to  $Q^2 > 5$ GeV$^2$) together with 
{\small BCDMS}~\cite{bb.bcdms} and {\small NMC}~\cite{bb.nmc} data, 
giving a good description of $F_2$ over the whole kinematic
range, surprisingly even when the fit is extrapolated at
the lowest $Q^2$ of $1.5$ GeV$^2$. From
the fit, the gluon density was extracted 
(fig.~\ref{fig.kin_h1}b), showing a sharp rise at low $x$.

More recently two further topics have been
adressed by H1 in this field and will be detailed in the
following: 
i) a dedicated  analysis of the very high $y$ 
region (up to $y=0.78$) 
of the 94 data 
used in conjunction with a NLO QCD fit
to the low $y$ part of the data, allowed the longitudinal structure function
$F_L$ to be  determined for the first time at HERA \cite{bb.h1fl}, giving a
consistent picture within the QCD framework;
ii) the analysis of $0.11 \pbinv$ data taken in 95  after 
the upgrade of the backward region of the H1 detector and 
with a shifted interaction vertex, therefore increasing the acceptance at
low $Q^2$ and low $x$. This allowed 
$F_2$ to be measured  down to $Q^2$ of $0.35$ GeV$^2$~\cite{bb.h1f295}, 
and  the
transition between perturbative and non-perturbative QCD to be reached.

\newpage
\begin{center}
{\large {\bf The Longitudinal Structure Function}}
\end{center}

In the single photon exchange approximation the relation between the
differential cross-section, the structure function $F_2$ and
the longitudinal structure function $F_L$ can be expressed as
$$ {d^2\sigma \over {dxdQ^2}}    =
    {{{2\pi\alpha^2} \over {xQ^4}} 
      \left[ (2(1-y)+y^2) F_2(x,Q^2) - y^2 F_L(x,Q^2)\right]} $$
where $F_L$ is related to the cross-section of transverse and the longitudinal 
polarised photons, $\sigma_T$ and $\sigma_L$:
$ F_L=F_2-2xF_1, R=\sigma_L / \sigma_T = F_L / (F_2-F_L $).\\
Note that the influence of $F_L$ to the DIS cross-section is most
important in the high $y$ region.

\begin{figure}[htb]
\begin{center}
\epsfig{file=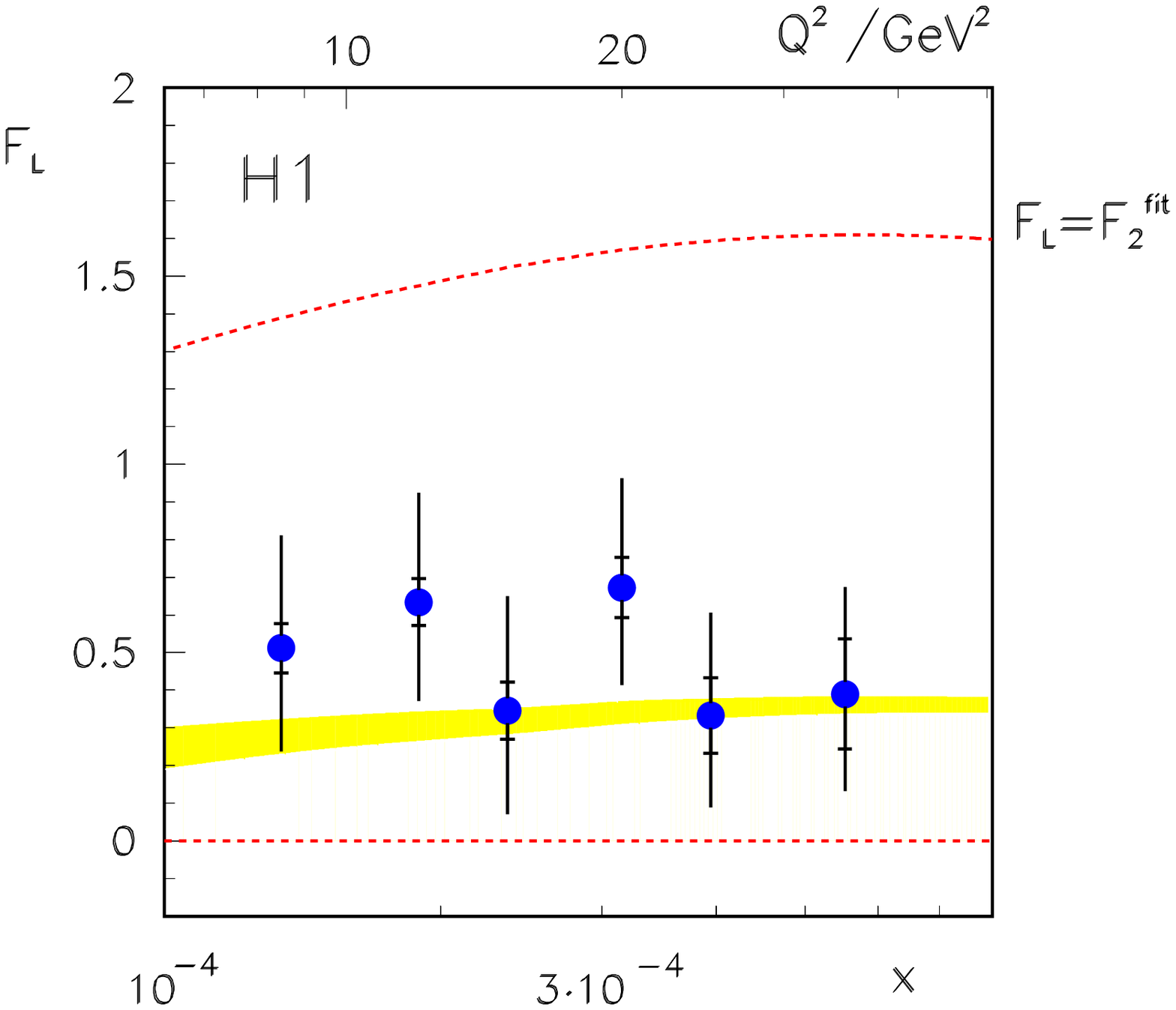,bbllx=60,bblly=210,bburx=545,bbury=630,width=8.cm,clip=}
\caption[]{\label{fig.fl}      
{\small \sl 
The longitudinal structure function $F_L$ at different values 
of $x$ and $Q^2$}}
\end{center}
\end{figure}

The conventional method to measure $F_L$ consists in unfolding 
the DIS cross-section measured at different center of mass energies 
as made by various fixed target experiments\cite{bb.flwhitlow,bb.flnmc}. 
The H1 collaboration has published a determination
of $F_L$ at $8.5$ GeV$^2 < Q^2 < 35$ GeV$^2$ and at very small $x$ values
between $1.3 \cdot 10^{-4}$ and $5.5 \cdot 10^{-4}$, 
using the so called ``subtraction
method'' \cite{bb.h1fl}:
as the behaviour of $F_2$ can be described with a good precision
by a NLO DGLAP fit over several orders of magnitude in $Q^2$,
such a fit is performed on the $F_2$ data at $y \leq 0.35$, where
the influence of $F_L$ is negligible. After an
extrapolation up to $y=0.78$ this contribution of $F_2$ to the cross-section
is subtracted and the difference attributed to $F_L$. The
results of this procedure is displayed in fig.~\ref{fig.fl} for six points
in $x$ and $Q^2$.

\vspace*{-0.5cm}

\vspace*{-0.5cm}
The $F_L$ obtained is consistent
with the Altarelli-Martinelli prediction \cite{bb.AltMart} obtained in 
perturbative  QCD (pQCD)  shown as the band in fig.~\ref{fig.fl}.
It excludes  the extreme limits of $F_L = 0$ and
$F_L = F_2$ by 2.3 and 4.0 times the total error on $F_L$

\newpage
\begin{center}
{\large {\bf $F_2^p$ at low $x$ and low $Q^2$}}
\end{center}

In the region of pQCD ($Q^2 \ge $ 1 GeV$^2$), 
the evolution of $F_2$ is  nicely
described by the model of Gl\"uck, Reya, Vogt (GRV) \cite{bb.GRV}, 
which evolves
valence like parton distribution from a low starting scale of $0.34$ GeV$^2$
according to NLO DGLAP equations. 
At $Q^2 \simeq 0$, in the photoproduction region, the measured cross-sections
are in agreement with the Donnachie-Landshoff model (DOLA) \cite{bb.DoLa},
which assumes a $Q^2$-independent Regge behaviour up to a few GeV$^2$,
due to the exchange of a ``soft'' pomeron.

\begin{figure}[htb] 
\begin{center}
\epsfig{file=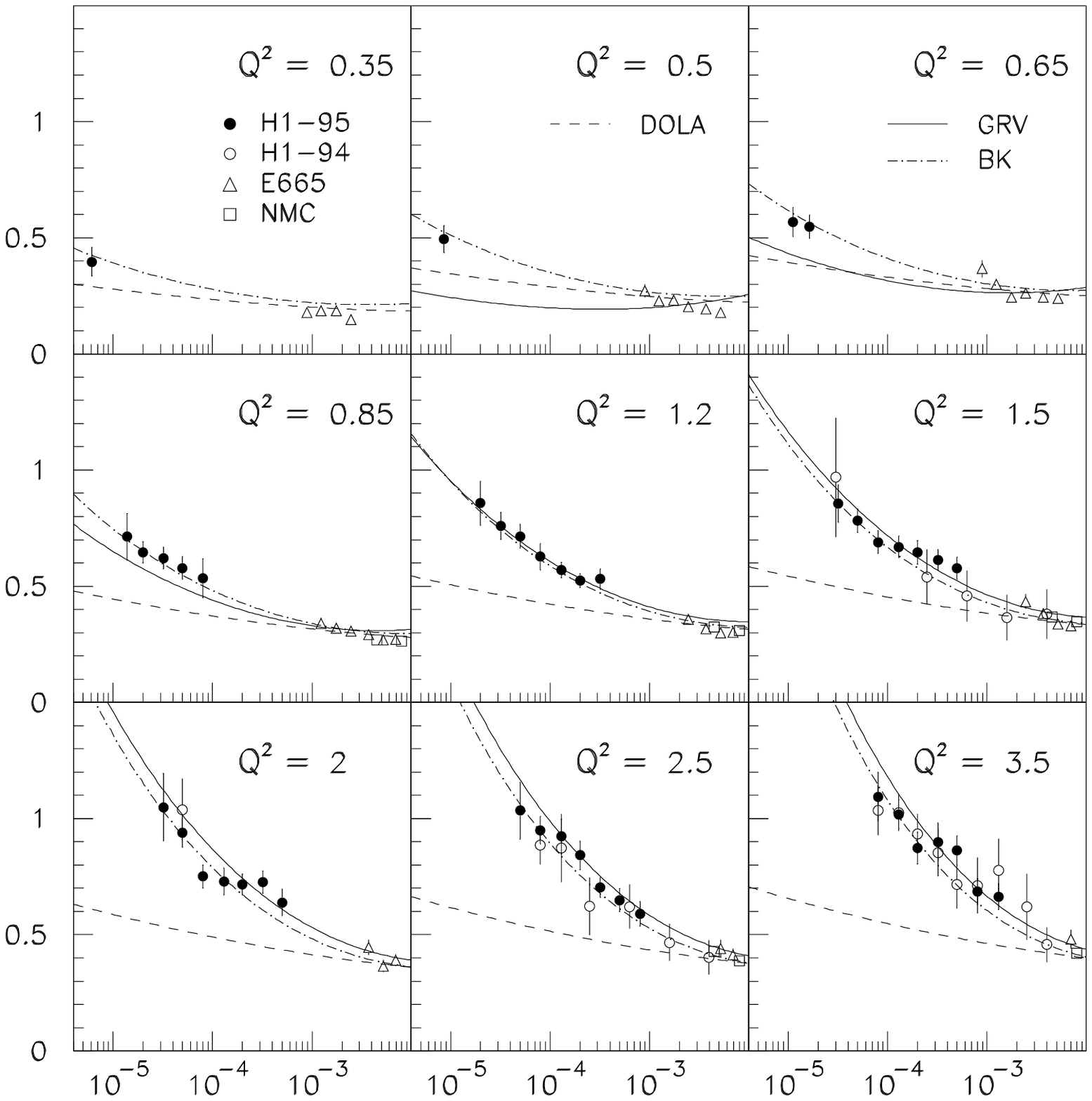,bbllx=55,bblly=180,bburx=520,
bbury=650,width=8.cm} 
\caption[]{\label{fig.f2lq}      
{\small \sl 
The $F_2$ structure function measurement at low $Q^2$ and low $x$ compared
to fixed targed data and to the GRV, BK and DOLA models.}}
\end{center}
\end{figure}

The recently published $F_2$ measurements at low $Q^2$  \cite{bb.h1f295} 
are shown in fig.~\ref{fig.f2lq}.
They  give new 
experimental information for the transition region between the perturbative
and non-perturbative QCD at $0.35$ GeV$^2 < Q^2 < 3.5$ GeV$^2$ and for
$x$ values down to $6\cdot 10^{-6}$, and are
compared to previous measurements done 
in ``fixed target'' mode  and to various phenomenological models.

One can notice the surprizing persistent rise of $F_2$ with decreasing $x$,
even at very low $Q^2$, although this rise gets weaker with decreasing $Q^2$.
Therefore the Donnachie-Landshoff model seems not to describe the data, even
in the lowest $Q^2$ bin. The weakening of the rise of $F_2$ is however 
too strong in the model of GRV for the data below $1$ GeV$^2$.
The Badelek-Kwiecinski model\cite{bb.BK} which combines
a generalised Vector Meson Dominance (VMD) approach at low $Q^2$ with a pQCD
one at high $Q^2$ is able to reproduce the $x$-$Q^2$ behaviour of $F_2$ in this
kinematic region.

\begin{figure}[htb] 
\begin{center}
\epsfig{file=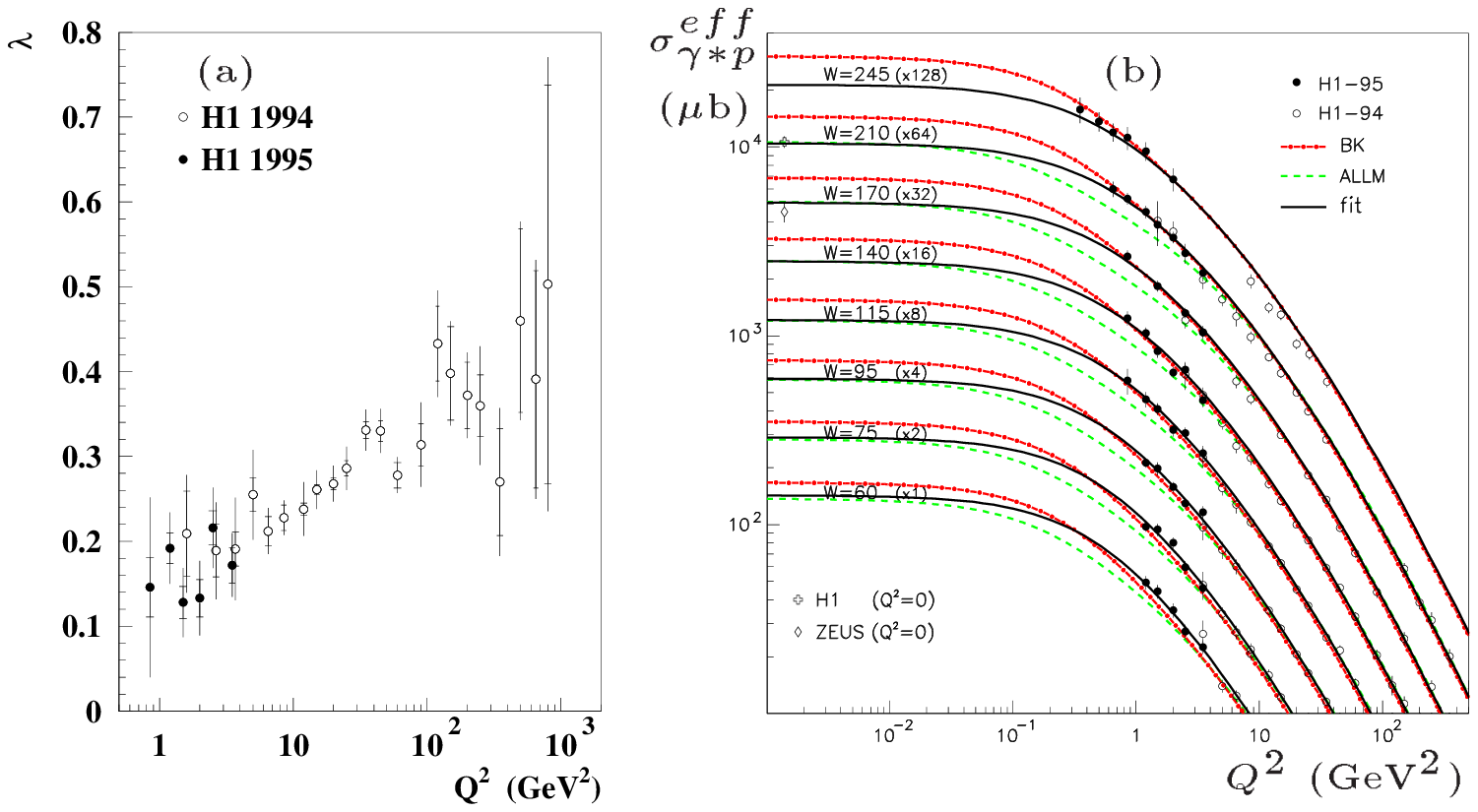,
bbllx=100,bblly=535,bburx=510,bbury=775,width=13cm}
\caption[]{\label{fig.weff}      
{\small \sl 
a) The variation of the exponent $\lambda$ expressing the slope of $F_2$ in 
$x$. b) Measuremens of the virtual photon-proton cross section
$\sigma^{eff}_{\gamma \ast p}$ as function of $Q^2$. }}
\end{center}
\end{figure}

The rise of $F_2$ as function of $x$ can be quantified by fitting 
$F_2 \propto x^{-\lambda}$ at fixed $Q^2$ values for $x < 0.1$. The values of
$\lambda$ for such a fit on the 94 and the 95 data are displayed in 
fig.~\ref{fig.weff}a. For $Q^2 \rightarrow 0$, $\lambda$  approaches
the value of  0.08 as expected in the photoproduction regime.

In fig.~\ref{fig.weff}b, the effective virtual photon-proton cross-section
is shown as function of $Q^2$ for different values of the invariant mass W.
We have:
$$
    {d^2\sigma \over {dxdQ^2}}    =
    {{{2\pi\alpha^2} \over {xQ^4}} 
      \left[ (2(1-y)+y^2) F_2(x,Q^2) - y^2 F_L(x,Q^2)\right]} 
      \equiv \Gamma\sigma^{eff}_{\gamma^* p}(x,y,Q^2)
$$
%
with
$\Gamma=\alpha(2-2y+y^2) / 2\pi Q^2x$.

Neither the BK model nor the model of ALLM \cite{bb.ALLM} 
can completely  describe the data
from $\simeq 0$ GeV$^2$ to high $Q^2$.
The model 
of ALLM gives a good limit for the measurement at $Q^2 = 0$ and at high 
$Q^2$, but tends to be too low in the transition region, whereas the BK model
is too high at $Q^2 = 0$. Therefore a fit was performed following the
BK approach with C$_{VMD}$ and $Q^2_{VMD}$ taken as free parameters, while they
are equal to $1$ and $0.8$ GeV$^2$ in the original BK model.
$$
F_2(x,Q^2)= C_{VMD} \cdot  F_2^{\rm{VMD}}(x,Q^2)+
{Q^2\over { Q^2_{VMD} +Q^2}} F_2^{\rm{QCD}}(\overline{x},Q^2+ 
 Q^2_{VMD})
$$
The resulting fit shown in fig.~\ref{fig.weff} gives
$C_{VMD} = 0.77$ and the unphysical value of $Q^2_{VMD}=0.45$GeV$^2$
($Q_{VMD}$ has to be larger than the mass of the mesons exchanged in the
VMD prescription).
Although it provides 
an overall good description of the data, the underlying
dynamics has still to be understood.

\section{Cross-Sections in DIS at High $Q^2$}
At high $Q^2$ the exchange of the $W$ or the $Z$ cannot be neglected 
anymore, and the differential cross-sections can be rewritten for the NC events
as  
    $$   \frac{d^{2}\sigma^{NC}(e^{\pm}p)}{d x d Q^2} \, = 
      \frac{2\pi\alpha^2}{xQ^4} \left(Y_+ { {\cal{F}}_2
      (x,Q^2)} 
    { \mp}  Y_- { x{\cal{F}}_3(x,Q^2)}\right) 
\hspace*{0.5cm}; \hspace*{0.5cm} 
    Y_{\pm}\equiv(1\pm(1-y)^2) $$
The $ { {\cal F}_2 } ,\; {{\cal F}_3}$ Structure
    functions  of the  proton include $Z^o$-propagator effects
and ew-couplings:
   ${\cal  F}_2$ contributes symmetrically for $e^+$ and $e^-$ scattering,
while
   ${\cal  F}_3$ is parity-violating, so its contribution changes
sign when $e+$ and $e-$ are exchanged.
For the CC events  the cross-sections can be expressed directly as a function
of the quark densities ($q=q(x,Q^2)$, with $q=u,d,s,c$), 
    $G_{\mu}$ being the  {Fermi--Coupling constant} and $M_W$ is the mass
of the $W$ boson in the propagator term.
    $$
    {\frac{d^2\sigma^{e^+p}}{dxdQ^2}} = 
    {\frac{G_\mu^2   }{\pi}} 
    {\frac{{ M_W^2}}{({ M_W^2}   + Q^2 )^{2} }} ( 
    {(\bar{u}+\bar{c})}  + 
    (1-y)^2 {(d+s)} )
    $$
    $$
    \frac{d^2\sigma^{e^-p}}{dxdQ^2} = 
    \frac{G_\mu^2 }{\pi} \frac{{ M_W^2}}{({ M_W^2}  + Q^2 )^{2} }
    ({({u}+{c})}  +  (1-y)^2 {(\bar{d}+\bar{s})})
    $$
Two analyses are presented in the following. The first, based on
6 pb$^{-1}$ of 1994-1995 data has been performed using the hadrons-only method,
in order to compare with minimal systematic bias the NC and CC differential
cross-sections~\cite{myriess}. 
The other using the full available statistics in H1
(14 pb$^{-1}$ of 1994-1996 data) and an optimized reconstruction was oriented
towards the study of very high $Q^2$ region~\cite{h1hiq296}.\\

\begin{figure} \begin{center}
      \epsfig {file=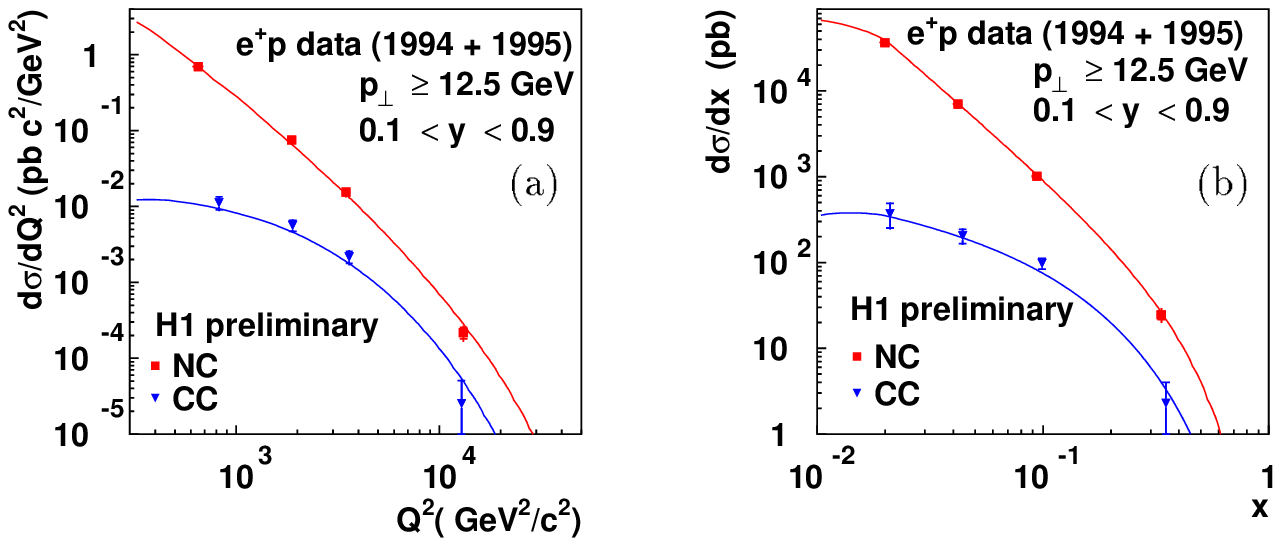,
bbllx=85,bblly=380,bburx=452,bbury=537,width=13cm}
\caption[]{\label{mycc} 
{\small \sl   
 NC and CC differential cross-section: a) $ \ {d\sigma}/{dQ^2} \ $; 
b) $ \ {d\sigma}/{dx} \ $.
}}
\end{center}
\end{figure}

\noindent $\bullet$
{\bf $d\sigma/dQ^2$ and  $ d\sigma/dx$ for NC and CC events} \\
In the first analysis, the total Cross Section  for Charged Current $e^+p$,
was obtained with a
a transverse momentum  cut $p_t>12.5 GeV$ and a kinematic range
restricted to $0.1<y<0.9$:
$${\rm \sigma_{tot}^{CC}  =  (25.2 \pm 2.5 \pm 0.8)  pb}
  \hspace*{0.5cm}    \mbox{\small {(preliminary)}}
$$
The differential cross-section $\frac{d\sigma}{dQ^2}$ (fig.~\ref{mycc}a), 
$\frac{d\sigma}{dx}$ (fig.~\ref{mycc}b) and 
$\frac{d\sigma}{dy}$ (cf~\cite{myriess}) show a good agreement with
the standard model on the full kinematic range available with that
statistics. The NC and CC cross-sections have comparable size for 
for  $Q^2 \approx m_Z^2$. In future
the $y$-differential  CC cross section is expected to provide
separate information on the  valence and  sea quark densities,
due to the different $y$ dependence of these two terms.\\

\noindent $\bullet$
{\bf The Very High $Q^2$ events} \\
With the increase of luminosity, the capabilities of the detector,
which was designed for the study of high $Q^2$ events, are fully 
exploited. Namely the H1 fine-grained Liquid Argon 
calorimeter ($\sim$ 44000 cells) allow the polar angle of the scattered
electron to be measured with a 2 to 5 mrad accuracy, its energy
with a resolution of   $\sigma(E)/E \simeq 12\%/\sqrt{E/\GeV} \oplus1\%$ 
and with an absolute scale known from test beams in most of the range
studied of $\pm3\%$.

\begin{figure}
  \begin{center}
\hspace*{-0.5cm}
\epsfig{file=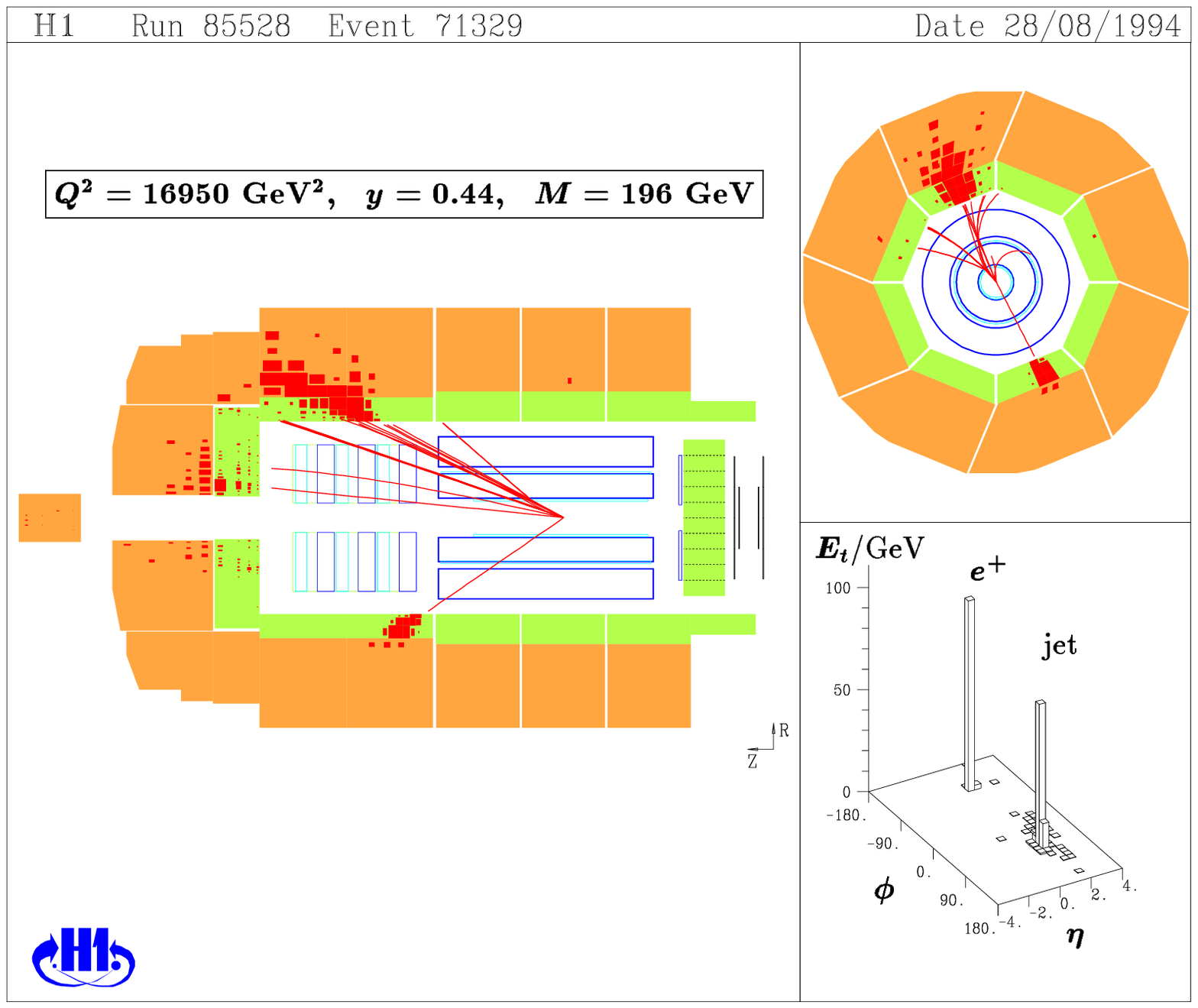,bbllx=90,bblly=200,bburx=500,bbury=540,
width=11cm,height=8.5cm}
\caption[]{\label{myNCdis} 
{\small \sl   
One of the 7 high $Q^2$ high Mass  NC events as visualized in the H1 detector
with its fine-grained Liquid Argon calorimeter. The positron is on the
bottom, the current jet on the upper part of the detector.
}}
\end{center}
\end{figure}

\begin{figure} \centering
\epsfig{file=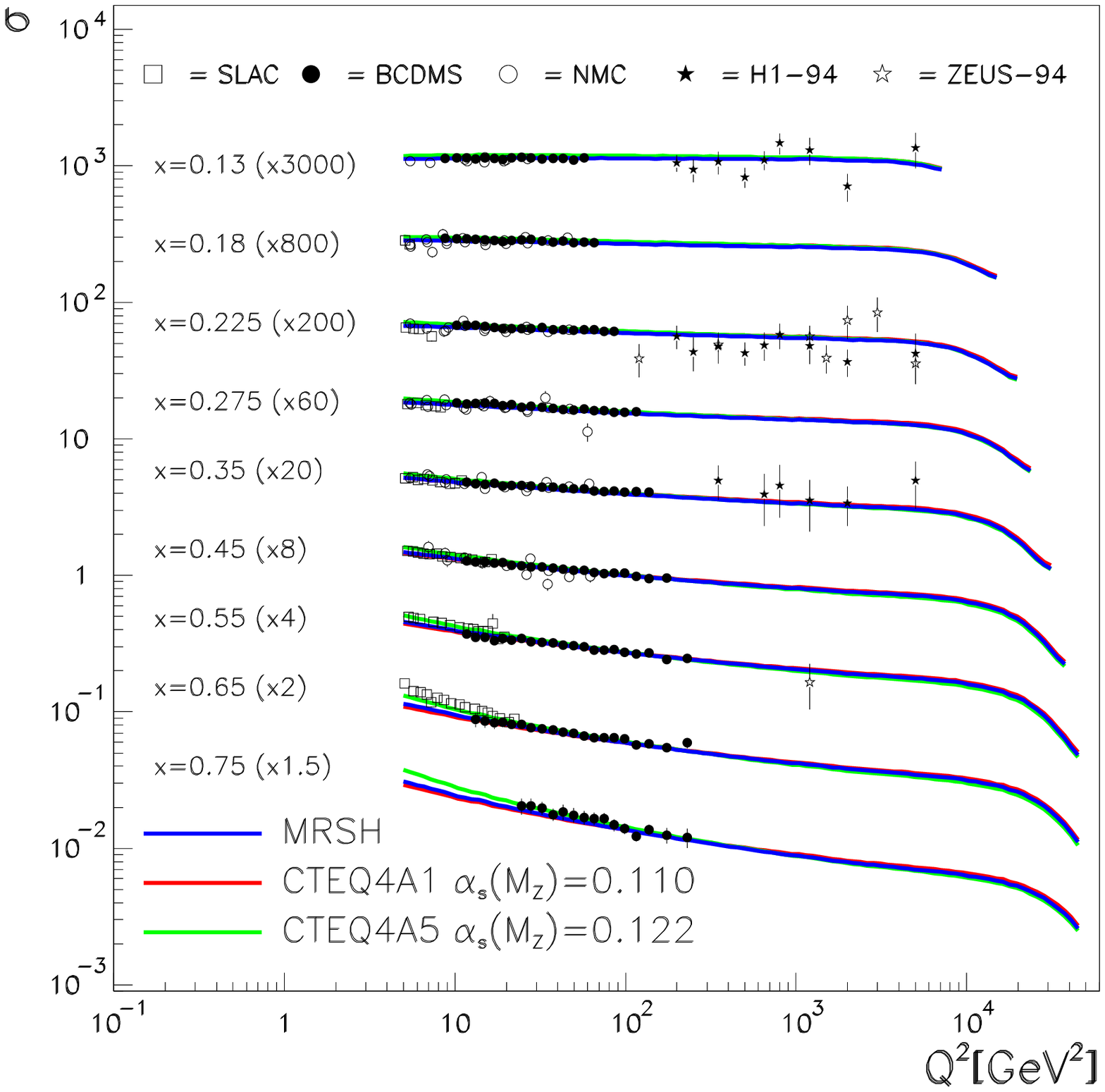,bbllx=40,bblly=150,bburx=540,bbury=650,width=8cm,
height=8cm}
\hfill
\epsfig{file=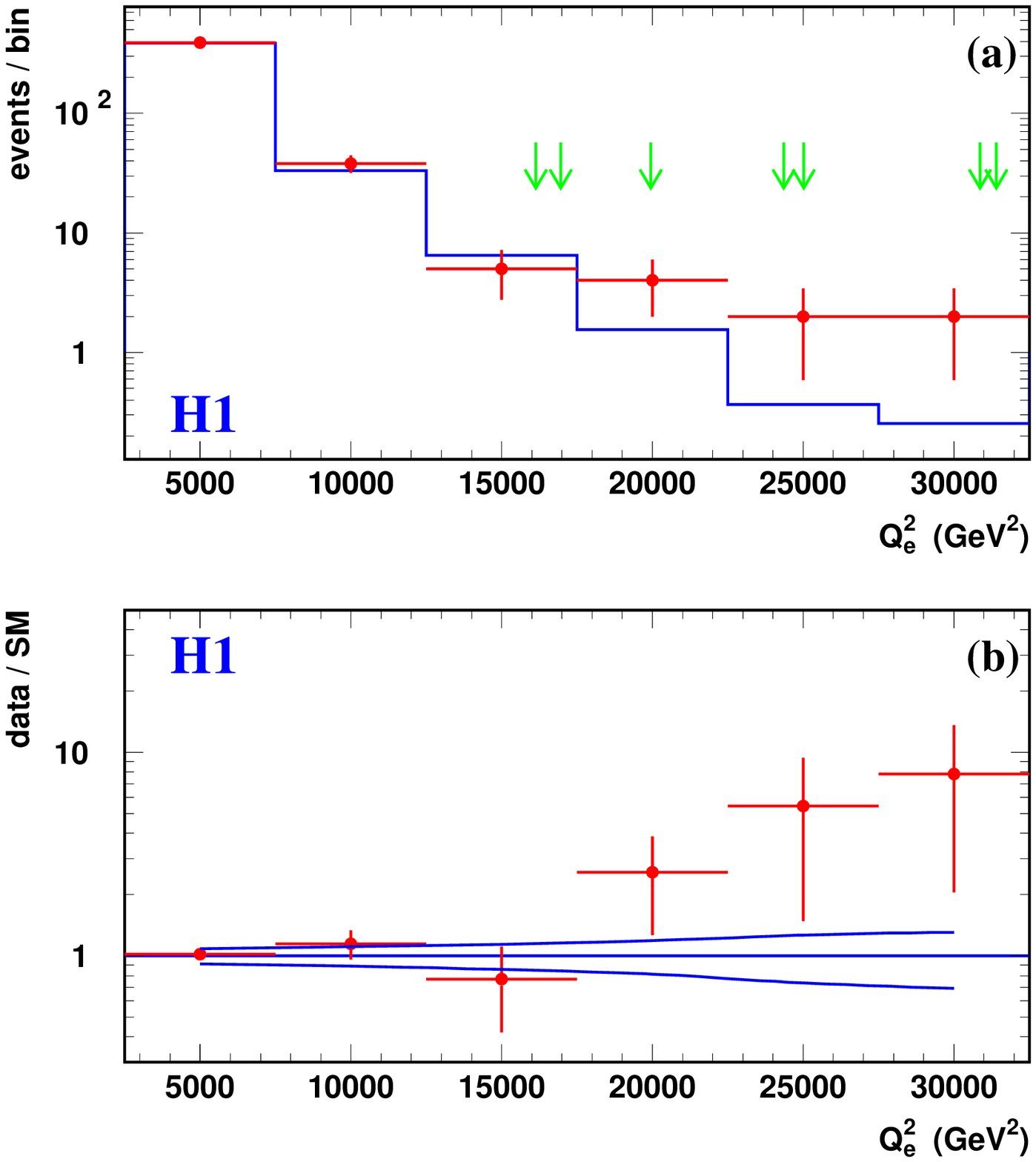,
bbllx=45,bblly=115,bburx=520,bbury=670,width=5.5cm,height=8cm}
\caption[]{\label{mycross}
{\small \sl Left:
Double differential cross-section 
  $ \rm  \frac{d^{2}\sigma^{NC}(e^{\pm}p)}{d x d Q^2}  $ as measured by DIS
experiments and HERA in 1994.
a) Raw $Q^2$ distribution for NC events displaying the excess
above 15000 GeV$^2$. The arrows point to the 7 
high mass, high $y$ events (see text). b)
ratio of data/standard model. The systematic error band is shown.
}} 
\end{figure}

The high $Q^2$ events are essentially background free
due to their striking signature (see fig.~\ref{myNCdis}).
The predictions of the standard DIS model are precise (see~\cite{bb.roberts}
for a more detailed review)
 in this kinematic
domain, since the electron probes the valence quarks at high $x$, which
are  constrained by the structure functions measured in the high statistics 
fixed target experiments.
The DGLAP evolution of these parton densities are also well understood at
NLO, and the theoretical error on the cross-section prediction is below 10\%,
including the uncertainties on $\alpha_S$, the shape of the ``input''
parton distributions, the evolution itself and the higher order QED
corrections.
These expectations have already been tested on the 1994 data
in the high $Q^2$ $F_2^p$
measurement at HERA, albeit with yet limited statistical precision, 
as can be seen in fig.~\ref{mycross}.

The 1994-1996 analysis is performed at $Q^2>2500$ GeV$^2$  both on
NC and CC. For the NC, the $e$ method is used as a reference, while a
complete cross-check has been done with the DA method.
The main selection cuts are:
{$ E_{T,e}>25\GeV$},
{$\Theta_e>10^o $},
{$0.1 < y_e < 0.9$},
plus additional conditions on the conservation of
transverse  and  longitudinal momentum.
The electron identification is done  using ``classic'' estimators
such as  shower shape, isolation in $\eta-\phi$, and a
loose track matching.
The remaining background after these cuts is negligible. 
For the CC the main condition is the requirement of a missing momentum
greater than 50 GeV.

The $Q^2$ distribution for NC events compared to the standard model histogram
can be seen in fig~\ref{mycross}a,b. An excess of events is visible above
15000 GeV$^2$, the 7 arrows pointing to the high $y$ events discussed below.
In CC events, no significant excess is observed
(4 events detected for 1.77$\pm0.87$ expected).
The significance of the  excess seen in NC events 
is found in the following table which gives
the probability of observing our number of events at  $Q^2>Q^2_{min}$.

\begin{small}
\begin{center}
  \begin{tabular}{|c|c|c|c|c|c|}
\hline
$Q^2_{min} (\GeV^2)$  & 2500   & 10000   &  { 15000}  & 20000 &30000\\
                                                            \hline \hline
$N_{obs}$             & 443     & 20     & { 12}     & 5 & 2     \\
                                                            \hline
$N_{DIS}$             & 426.7  $\pm 38.4$   & 18.3  $\pm 2.4$
  & { 4.71}  $ \pm 0.76$  & 1.32  $\pm 0.27$ & 0.23 $\pm 0.05$ \\
\hline
${\cal{P}}(N \geq N_{obs})$
  &  0.35  & 0.39   &  $6 \ 10^{-3}$    & $1.4 \ 10^{-2}$ & $2.3 \ 10^{-2}$ \\
 \hline 
\end{tabular}
\end{center}
\end{small}

If applying a cut $y>0.4$ in order to remove the largest part
of the standard DIS expected events,  7 events at $Q^2>15000$ GeV$^2$ remain
(see arrows in fig.~\ref{mycross}a) and they
 are clustered  at an invariant  mass of $200~\pm12.5$ GeV, where only
$0.95\pm0.18$ are expected. Several other methods have been used
to derive the mass and all give the average result  of 200 GeV (within
2 GeV)~\cite{h1hiq296}. The $\Sigma$ method has also been applied 
since it is independent of initial state QED radiation, and the results
are given in table 2. The only event which might be radiative (event 5)
as can be seen by comparing the kinematic variables reconstructed by
the  $e$, DA and $\Sigma$ methods,
is showing basically no change in its reconstructed  mass.

\begin{small}
\begin{center}
\begin{tabular}{|c|c|c|c|c|c|c|c|c|c|}
\hline
evt & ${ M_e}$  &  $M_{\Sigma}$ &  $M_{DA}$ &
 $y_e$  & $y_\Sigma$ & $y_{DA}$ &
 $Q^2_e$  & $Q^2_\Sigma$ & $Q^2_{DA}$ \\
& (GeV)& (GeV) & (GeV) & & & & (GeV$^2)$ & (GeV$^2$) & (GeV$^2$) \\
\hline 
1& 196  & 196 & 198 & .439  & .443 & .434 & 16950  & 17100 & 17100\\
2& 208  & 209 & 200 & .563  & .592 & .582 & 24350  & 25930 & 23320\\
3& 188  & 188 & 185 & .566  & .561 & .573 & 19950  & 19760 & 19640\\
4& 198  & 196 & 199 & .790  & .786 & .787 & 30870  & 30230 & 31320\\
5& 211  & 210 & 227 & .562  & .525 & .526 & 25030  & 23120 & 27100\\
6& 192  & 190 & 190 & .440  & .501 & .443 & 16130  & 18140 & 16050\\
7& 200  & 202 & 213 & .783  & .786 & .762 & 31420  & 31940 & 34450\\
\hline
\end{tabular}
\end{center}
\end{small}

In conclusion the accumulation of these events at $M=200$ GeV
is genuine, but is not confirmed by the  
distribution of the ZEUS events at high $Q^2$ and high
$y$. However
ZEUS has also an excess of a few events in NC at very high $Q^2$.
The systematic effects being well under control, both on the expectation
side as on the detection side,
more data are  needed to understand if the observed excess at
high $Q^2$ is  due
to a statistical fluctuations or to signs of new physics.

\begin{center}
{\bf SUMMARY}
\end{center}
A selected sample of H1 results  presented at this workshop 
in the different areas of DIS has been summarized. The increase of the 
luminosity will bear significant improvements in all these fields, showing the
potential of the HERA collider. Most awaited is an answer on the high $Q^2$
puzzle which might reveal signs of physics beyond the Standard Model.

\vspace*{1cm}
\begin{center}
{\bf ACKNOWLEDGMENTS}
\end{center}

I would like to thank the organizers for an exciting and enjoyable
workshop. Thanks also to the 15 H1 speakers of the parallel
sessions who provided me with much of the material shown here
and to John Dainton for a careful reading of the manuscript. Many thanks
to Ursula Bassler for her great help both in the preparation
of the talk and in the completion of the written report.

\end{document}